# The historical development of X-ray Absorption Fine Spectroscopy and of its applications to Materials Science

Annibale Mottana

Department of Geosciences, University of Roma Tre, Italy

e–mail: annibale.mottana@uniroma3.it

Augusto Marcelli

Istituto Nazionale di Fisica Nucleare, Laboratori Nazionali di Frascati, Italy

e–mail: augusto.marcelli@infn.lnf.it

**Abstract.** This essay sketches the development of X-ray Absorption Fine Spectroscopy (XAFS) ever since the second half of 20[th] century. At that time, synchrotrons started competing with X-ray discharge tubes as the sources of the excitation able to show the pre- and near-edge structures (XANES) and extended oscillations (EXAFS) that characterize the X-ray absorption edge of solid matter. Actually, modern XAFS began after 1975, when the hard-X-ray synchrotron radiation derived from storage rings took over. Ever since, XAFS greatly contributed to both technical refinement and to theoretical development of Materials Science. Although a unified theory of X-ray fine absorption has not been reached yet, many XAFS advancements benefited of theoretical models and complex calculations made possible by the continuous growth of the computing power, while contributing to developing new or previously never used materials.

## 1 Introduction

Wilhelm Conrad Röntgen (March 28, 1845 – February 10, 1923) detected absorption of X-rays by matter on the very evening he discovered X-rays (November 8, 1895). Some days later, he also realized that different materials display different degrees of transparency, as he took

> an image of his wife's hand which showed the shadows thrown by the bones of her hand and that of a ring she was wearing, surrounded by the penumbra of the flesh, which was more permeable to the rays and therefore threw a fainter shadow.[1]

His first, simple observations were crucial not only for the discovery of X-rays but also for their quick expansion throughout the world. It is well known, unfortunately, that Science is considered by most people to be of little use ("pure") if does not produce practical results at once. The transparency noticed by Röntgen found a practical application almost immediately in clinical Medicine: numberless clinical transparency images were taken, making Radiology the most productive diagnostic method to date.[2]

Soon X-rays were also tested in detail for scientific reasons. However, reviewing all branches of Science that benefited from X-rays is not the aim of this contribution. Rather, we will describe the development of a field of X-ray-based research on solid matter that started over half a century after Röntgen's discovery: X-ray Absorption Fine Spectroscopy (XAFS).

XAFS is the modern development of X-ray Absorption Spectroscopy (XAS), a branch of X-ray-based Physics that began over ten years later than Röntgen's X-ray discovery, and a couple of years later than X-ray Diffraction (XRD), the first widespread form of application of X-rays for scientific purposes. For a long time XAS was poorly considered or even under-esteemed, overshadowed as it was by XRD extraordinary success. Indeed, XRD had made solid-state research take off because it could determine the lattice structure of ordered matter, but it could nothing when such an order was not present, as in amorphous materials, highly disordered systems and liquid solutions. This was indeed XAS' deed, but this took long time to attain be understood and even more to attain successful results. Only during the second half of last century XAS took off too, and in the new form of XAFS. In order to reach the front-line of Science, it had to wait for the availability of powerful and brilliant X-ray sources such as the electron synchrotrons first, and then the much more stable and brilliant storage rings. In addition, the availability of powerful computers and advanced computational methods pushed to better insight into theory (an essential part of Science), by allowing simulations that would both validate and crosscheck the results of experiments.

---

[1] This sentence is in the autobiography Röntgen wrote at the time of the Nobel award, later translated into English and published in the *Nobel Lectures* (1967).

[2] As early as on February 3, 1896, at Dartmouth Medical School and Hospital, Hanover, NH, USA, the first photographic image of the fractured wrist of a boy was taken (Spiegel 1995 p 242) and the injured bone could be appropriately compounded: first clinical application of X-rays in the whole world.

Now over 50% of the requests for time reaching synchrotron laboratories from industrial customers involve one or another X-ray absorption method, making XAFS the choice field for any study on innovative materials. XAFS is also the best tool to evaluate their technological applications, particularly for Chemistry, which is the science that creates new materials; for Biology to characterize biomolecules, both natural and synthetic; finally, yet just as importantly, for the Earth's scientists, to characterize amorphous or even glassy materials, where XRD is of little help, if any.

The historical development of XAFS has been reviewed several times, but only by bit and pieces. Actually, there is one authoritative general review (Stumm von Bordwehr 1989), which however stops in the year 1975, when synchrotrons started replacing discharge tubes as the X-ray sources. In addition, there are short review papers published in the proceedings of the many international conferences taking place ever since 1982 (e.g., the series "EXAFS and Near Edge Structure", or "Advances in X-ray Spectroscopy", etc.), and there are also accounts by or about people who contributed to work in progress (e.g., Citrin 1986; Doniach et al. 1997; Lytle 1999; Stern 2001; Mottana 2003; etc.). Nevertheless, there is no appraisal yet of the pros and cons, which Materials Science has derived from XAFS during fifty odd years of life and development. This is indeed the purpose of our still limited (in terms of space) review.

## 1.1. How XAS came into being

1913 was a most extraordinary year for Physics. It saw the birth of some aspects of this science that are basic for its modern state-of-art, such as the first formulation of the atomic theory (Niels Bohr); the first crystal structure solution by XRD (W. Henry Bragg and W. Lawrence Bragg); the essentials of X-ray fluorescence (XRF) analysis (Henry G.J. Moseley). It was the year too when [August] Julius Hedwig (October 17, 1879 – January 26, 1936) in Germany and [Louis-César-Victor-] Maurice de Broglie (April 27, 1875 – July 14, 1960) in France, working independently but both using a modified type of spectrometer in a reverse way to impinge a metal sheet with monochromatic X-rays, detected on photographic plates some sharp or diffuse lines and bands that appeared next to the strong, enhanced edge of the investigated metal atom (platinum or tungsten) in the region of shorter wavelengths: a spectrum, then. On this rather weak basis, M. de Broglie (1913) dared proposing a structure for X-rays that was similar to that - long since known - of light spectra. Neither Herweg nor de Broglie were completely confident in their observations[3]. The former scientist (whom priority should be assigned, as he submitted his paper on June 30) gave up with this kind of research. The latter one (who presented his first note on November 17) was heavily criticized and even had to admit he had misinterpreted his first photos. Nevertheless, he kept on experimenting and communicating new results for some years, thus amply justifying the reputation of him as the man who "*invented X-ray spectroscopy*" (Lytle 1991 p 123). The existence of modulations at the X-ray absorption edge of metals was first established without doubts between 1918 and 1920 by [Karl] Wilhelm Stenström (January 28, 1891 – November 7, 1973) and Hugo Fricke (August 15, 1892 – November 10, 1972), both working at Lund University under the supervision of [Karl] Manne [Georg] Siegbahn (December 3, 1886 – September 26, 1978), who earned the 1924 Nobel Prize because of it.

Actually, XAS experienced a great upgrade at the beginning of the second decade of 20$^{th}$ century, when it was instrumental in probing the electronic structure of the atoms and in the successful development of quantum theory. Many people contributed to this early phase of basic research, and by the final years 1920s, the fundamentals of XAS were firmly set. Then, limitations inherent in the X-ray sources and detectors restricted its further development for many years especially on what the method was most in the need for: the interpretation of the fine structures detectable around the absorption threshold (known, at that time, as the "*Kossel structure*"). They had been understood in their generality, but not really worked out in detail. Moreover, other features at higher energy (known as the "*Kronig structure*") had been detected, studied and explained, however without reaching real consensus on their meaning among scientists, mainly because experimental results and calculated data rarely matched.

---

[3] *Eine Aneinanderreihung von Flecken, die verschiedenen Wellenlängen entsprechen* (Herweg 1913 p 556); *un véritable spectre de raies, ayant tout à fait l'aspect des photographies de spectre lumineux, avec des raies fines or diffuses, des bandes, etc*. (de Broglie 1913 p 925).

For almost 30 years, it was clear that in order to really reach the front-line if Science XAS needed more intense X-ray sources, much better detection methods, and - last but not least - a better theoretical background than the theories formulated by Walther Kossel (1920) and Ralph deLaer Kronig (1931, 1932a,b).

## 2 XAFS moves to synchrotron sources

### 2.1 Improvements in the apparatus

A first upgrade in the experimental part of XAS studies that would eventually lead to XAFS occurred only after the Second World War (WWII), when Norelco designed a novel type of diffractometer for powder XRD use. Indeed, a Norelco vertical powder diffractometer was the instrument modified by Robert Van Nostrand (November 28, 1918 – January 20, 2012) to measure the absorption coefficient of the catalysts produced by the company he was employed in. As monochromator he used a single crystal of Si or quartz set at the centre of the Rowland circle, i.e., where the sample to be analysed by XRD is, and he recorded the reflected beam intensity using a counter that could be rotated stepwise. He first was doing it for the beam passing free ($I_0$), than for the beam filtered by a finely powdered sample ($I_1$). The absorbance, $\ln(I_0/I_1)$, was plotted as a function of θ, the rotation angle of the diffractometer, the intensity at each θ angle being a fraction of the maximum occurring at the absorption jump, which he normalized to one irrespectively of the operation conditions, thus introducing a normalized absorption coefficient (Van Nostrand 1960). In this way, which mimics most of the powder-XRD practice, he introduced into XAS the manner of presenting data that has now become standard[4] not

---

[4] *For the study of the chemistry of catalysts and other non-crystalline systems this technique may have a role comparable to that of X-ray and electron diffraction in crystalline systems* (Van Nostrand 1960 p 184). Indeed, he was wrong only in that

only in the industry but also among scientists. Indeed, he could use the absorption data not only for internal comparison, but also to exchange with outside laboratories, thus creating the first XAS data bank. Despite this substantial improvement, which prompted not only other experimental studies but also reconsideration of the entire theory (see later), nothing highly significant would have really happened with XAS had a new intense and brilliant source of radiation not entered physical sciences: the synchrotron.

Particle accelerators had been studied theoretically during WWII and were built for research in the early years 1950s. Using these early accelerators the first $K$-edge absorption spectrum ever-recorded was that of Be metal (Johnston and Tomboulian 1954: Fig. 1) using the soft X-rays emitted by the Cornell synchrotron. In Europe, the first ones were the Al and Cu $K$-edges recorded in May 1963 at the Frascati electron synchrotron (Cauchois et al. 1963; Mottana and Marcelli 2013).

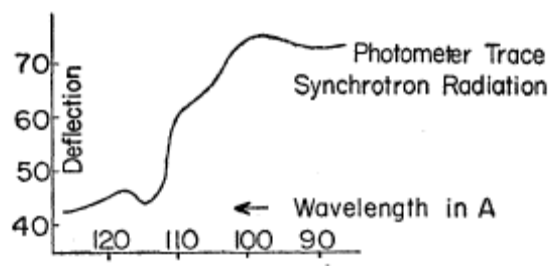

Fig. 1. The first absorption spectrum recorded using synchrotron radiation: the B K-edge using the 300 MeV electron synchrotron (after Johnston & Tomboulian, Phys. Rev. **94**, 1954 - p. 1589 Fig. 5, strongly reduced).

The electron synchrotrons, despite their limited energy (~1 GeV) and large instabilities, made detailed studies in the soft X-ray energy range possible. Actually, such studies were few (e.g., Sagawa et al. 1966; Jaeglé and Missoni 1966; Balzarotti et al. 1970), and were limited to the edges of gases and of certain metal atoms. Yet, they promoted not only the use of

---

he did not foresee that the technique could be applied just as successfully to natural samples that are significant for the oil and mining industries, and more recently to such environmental problems such as pollution by dust and aerosols.

synchrotron radiation (SR), but also the reactivation of studying XAS basic theory. Unfortunately, it is a historical fact that, in spite of all advantages offered by orbit-derived radiation (Parrat 1959), SR as the X-ray source did not become really popular among the XAS community until storage rings were introduced i.e., 10-15 years after installing the first electron synchrotrons. Such a long time delay had several good reasons: the conventional source was much cheaper, and had the advantage of being stable and reliable, besides being at home.

The approach to Science changed only in the middle of the years '70s, and compelled people to move from their laboratories to large facilities operating stable storage rings i.e., to radiation sources much more suitable for high-quality research. Their highly increased brilliance, combined with the production of hard X-rays, made it possible to measure the *K*-edges of heavy metal atoms, and these in turn widened the use of XAFS to branches of Science well outside Physics, such as Chemistry, Geochemistry, Biophysics, Metallurgy, etc.

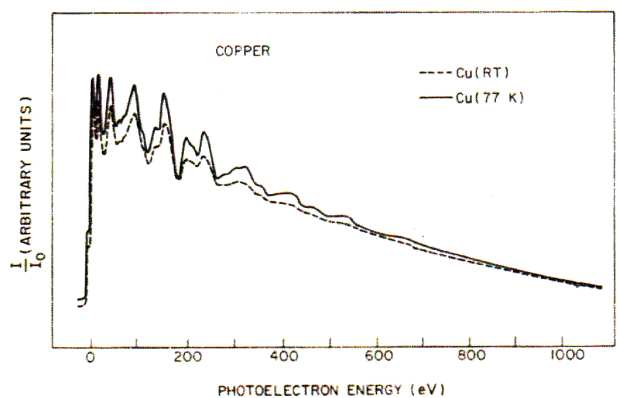

*Fig. 2. The first XAS spectrum recorded using the synchrotron radiation extracted from the SPEAR I storage ring (Eisenberg et al., 1974 - p 806 Fig 1).*

The first experimental spectrum that used SR extracted from a storage ring by a bending magnet was recorded on metallic copper at the Stanford Synchrotron Radiation Facility (SSRF, now SSRL; cf. Doniach et al.

1997), the orbiting storage ring being SPEAR I (Fig. 2).[5] Irrespectively on whether those experiments were done at room or liquid nitrogen temperature, the improvement in intensity over a conventional source was beyond any expectation. It was calculated to be $5 \times 10^4$, together with a $10^4$ enhancement in the signal-to-noise ratio, so that the recording time of a complete EXAFS spectrum could be reduced to 30 min (Kincaid and Eisenberger 1975 p 1361). The following statement by one of the founding fathers of modern XAFS is worth quoting, as it marks the change of paradigm: *In one trip to the synchrotron we collected more and better data in three days than in the previous ten years. I shut down all three X-ray spectrometers in the Boeing laboratory. A new era had arrived* (Lytle 1999 p 132).

Over a short period of time substantial improvements were conceived and put into operation aiming to record even better spectra, not only by new techniques such as fluorescence detection based on solid-state detectors and optimized for diluted systems (Jaklevic et al. 1977), but also by an increase of the SR flux that bending magnets extracted from the source. Wiggler magnets were developed jointly in the early years '80s, and first used at SSRL (Winick et al. 1981), then at Frascati National Laboratories (LNF), where the ADONE storage ring had started operation in 1978. They proved to be immensely useful because they produced a SR spectrum with a much higher critical energy, thus providing a high-enough intensity for all absorption experiments requiring high-energy photons (> 20 keV). Consequently, many beam lines optimized for spectroscopy moved from bending magnets to wigglers devices. Further energy gains were obtained by developing undulators, which increase power and brilliance of several orders of magnitude.

Because of all this upgrade, modern XAFS blossomed, and a great number of data became available to renew and test the theory, which was largely lying still on Kronig's one (1931). Nevertheless, there were other technical limitations in the apparatus and particularly in the computing power. They made so that two specialized techniques developed independently, in both experiment and calculation: (a) EXAFS, which provides a wealth of quantitative bond distance information based on single scattering processes, and (b) XANES, which implements the quantitative structural information deduced from EXAFS with information on the electronic properties, and also recognizes the local coordination and gives bond angles information. The time-maps show that the developments of these two techniques were not the same despite both used the same experimental apparatus. EXAFS progressed rapidly; XANES lagged behind hampered by its own complexity. The use of new acronyms underlines a new theoretical and practical approach. In fact, XAFS main purpose shifted towards developing a structure-determination technique that would stand side-by-side or even compete with the XRD-dependent science of Crystallography, by then mature and able to determine the average crystal structure with accuracies in bond lengths ±0.001 Å and angles ±0.1°. During the years '80s, it was clearly understood that determining the atomic structure with all its local defects is the reason behind any study on solid-state matter, as number and distribution of defects determine and/or modify physical properties. Such a result can be best attained by XAFS, which is a local probe, and is reached even better when its results are combined with information on the average crystal structure. The change of paradigm of Materials Science was complete, and XAFS was finally in the front line of Science.

---

[5] The same paper (Eisenberg et al., 1974) shows in addition the Cu XANES spectrum recorded on a Cu-porphyrin: this is the first spectrum of an organic molecule ever recorded.

For a rather long period of time scientists working with XAFS could make use of SR from storage rings only parasitically, but the evident rapid progress they made and the usefulness of the technique for fundamental research and industrial applications suggested first creating dedicated beam lines, then entirely dedicated laboratories. Intensity and tunability were the properties most required and best made use of. First generation storage rings had been designed for researches on particle physics, but were implemented with partly dedicated or parasitic beam time and with beam lines especially designed for XAFS already in the early '70s. Second generation storage rings designed to run as fully dedicated sources started operation around 1975 (the first one being SOR in Tokyo, Japan, a 380 MeV ring). There is now plenty of second- and third-generation storage rings in operation, where all types of XAFS experiments can be performed. The largest, most efficient and best-equipped facilities are ESRF at

Grenoble in Europe (6 GeV), APS II at Argonne in USA (7 GeV) and SPRING-8 in Japan (8 GeV).[6]

The continuous upgrade of the synchrotron radiation facilities, such as the increased current and stability and, more recently, the availability of topping up modes, may explain the continuous increase in number and quality of XAFS studies, which now involve all possible kinds of materials and phenomena, and operate even under extreme P,T conditions and high magnetic fields. To achieve these results, and in particular to perform experiments on extremely diluted systems, the availability of new solid-state multi-element detectors was absolutely needed. Computer Science also contributed with new methods that make the old cumbersome ones simply legendary. The extensive use of computers is also responsible of the software packages now available for EXAFS and XANES analyses.

## 2.2 Development of theory

R. Stumm von Bordwehr (1989) closed his review by reaffirming (p 444) why he decided to finish his work with the state-of-art of XAS in 1975: *at this date the theory of XAFS was established ... and* [the year 1975] *marks the beginning of an explosive growth in the field* (p 379). In fact, the change from XAS to XAFS and all related growth were propelled not only by the introduction of synchrotron as the radiation source of choice, but because the obsolete Kronig's theory was profoundly revised and implemented. The poor calculations resulting from this theory forced researchers to develop new methods and strategies, and a virtuous loop started of experimental data requiring theoretical refining.

For clarity, we anticipate that modern XAFS theory describes the entire absorption spectrum as resulting from the scattering of a photoelectron, which progressively damps out in energy moving from a multiple-scattering (MS) regime, which entails many scattering contributions generated by the photoelectron with neighbouring atoms around the photoabsorber, containing plenty of information on the local geometrical structure, i.e. bond distances and bond angles, to a single-scattering (SS) regime, which involves only couples of neighbouring atoms and contains information on bond distances (Fig. 3).

---

[6] To date, there are 67 synchrotron radiation facilities (www.lightsources.org/) in operation across the world, with around 152 beam lines optimized for XAFS measurements: 61 in the Americas, 54 in Europe and 37 in Asia (H. Oyanagi, personal communication).

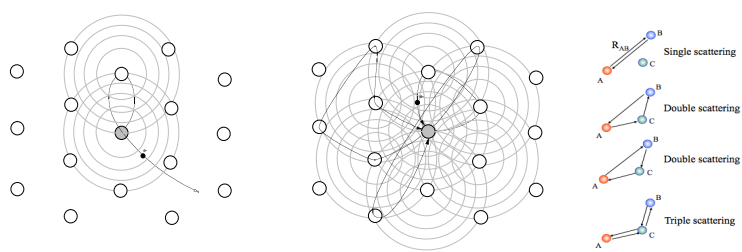

*Fig. 3. Pictorial view of the photoelectron scattering processes: (left) single scattering by a single neighbour atom = EXAFS; (right) simple multiple scattering pathways associated to processes generated by interactions with neighbouring atoms set at different distances and angles = XANES (centre). All events generate waves and contribute to the absorption cross section, before the photoabsorber atom returns to its fundamental state in ca. $10^{-15}$ s.*

**3.2.a. EXAFS Spectroscopy**. In his recollection of the EXAFS first steps (1999), Farrel W. Lytle (November 10,1934) makes a strong point at stating that, before going to SSRL, *all the data had been obtained with the conventional X-ray sources in our laboratory* (p. 132). Moreover, he stresses that the new theory of EXAFS[7] was conceived by him and first presented as nothing more than a preliminary hypothesis (Lytle 1965) because it was totally different (*a break*: p 129) with respect to Kronig's theory, which at that time was still the reference one, although it was 35 years old. Lytle gradually improved his theory by working in close cooperation with Dale E. Sayers (1940-2004) and Edward A. Stern (1930) (Sayers et al. 1970, 1971, 1972). Actually, it is clear that he had felt the need of developing a new theory because he had a wealth of experimental data gathered using a Siemens automatic diffractometer with conventional X-ray source that he had to justify to his company heads. He had modified his instrument, which operated in a horizontal arrangement, essentially mimicking the vertical spectrometer set up by Van Nostrand. To Lytle's luck, ten years later his horizontal orientation turned out to fit perfectly the requirements of storage rings, so that the entire EXAFS technique his group had developed for home studies was quickly transferred to the newly available, powerful and time-sparing SR sources.[8] Additional work (especially for the correct use of the Fourier transform) was required to perfect the quantitative interpretation technique and this work too the three scientists performed jointly. The final EXAFS package covered all three major points: theory (Stern 1974), experiment (Lytle et al. 1975), and determination of structural parameters (Stern et al. 1975). The complete EXAFS formula was:

$$\chi(k) = -\sum_j \frac{N_j}{kR_j^2} \left| f^j(k,\pi) \right| \exp\frac{-2R_j}{\lambda} \sin\left[2kR_j + \delta_j(k)\right] \exp\left(-2k^2\sigma_j^2\right)$$

---

[7] This acronym is credited explicitly by him to J.A. Prins (Lytle 1999 p 130).

It may appear strange that EXAFS (formerly known as a weak "secondary absorption" and noticed several years later than the "fine structures" near the absorption edge; cf. Mottana, in preparation) was clarified earlier and quantitatively solved more deeply than the "fine-structures" near the absorption edge, known from 1918-20. This historical fact can be easily explained. EXAFS structures had been disregarded by early researchers essentially because of the limitations inherent in their photographic detectors, but they could be recorded well by the counters of modified diffractometer, which also span over much wider angles. Moreover, both Van Nostrand and Lytle started working on this region of the XAS spectrum because they looked for few diagnostic features that could be usable for practical purposes, and simple enough as to require only a simple mathematical analysis.

The theoretical innovation that Lytle's group proposed for EXAFS is based, in fact, on phenomenological rather than abstract theoretical grounds. It considers the structures in the EXAFS region as due to interference of electrons: when an X-ray photon impinges an atom, it excites a core electron, which moves away through the atomic lattice as a photoelectron having energy equal to that of the incoming photon minus the core binding energy. Alternatively, the photoelectron may be seen to propagate through the lattice as a spherical wave, its wavelength decreasing with increasing the energy of the impinging photon. This spherical wave interacts with waves arising from the atoms neighbouring the absorber within a certain distance, and each atoms reacts behaving as a point-scatterer, so that the propagating total wave is the sum of all scattered waves. The total absorption (and absorption coefficient) is determined by the dipole transition matrix element between the initial core state and the photoelectron final state, which in turn is determined by the superimposition of the outgoing and incoming spherical waves. Inter-atomic distance and phase relationships between these two waves determine final state amplitude and modulation of the effects; the frequency of the damped oscillations in the spectrum depends on the distance between absorber and back-scatterer atoms; the amplitude is proportional to the number of the back-scatterer atoms, i.e. to the coordination around the absorber. EXAFS is indeed a local probe i.e., it is insensitive to the Long Range Order (LRO), as the photoelectron has a mean-free-path typically less than 10 Å (Müller et al. 1982). Nevertheless, EXAFS is a powerful method to study the Short Range Order (SRO) and is chemically selective. In fact, the large energy separation between different inner shells makes it possible to tune the incident X-ray photon energy of a particular core level corresponding to the chosen atom edge; the spectrum is then scanned up to *c*. 1000÷2000 eV above edge, provided there are no interfering contributions by any edges of other atoms present in the sample.

---

[8] *Synchrotron radiation revolutioned the experimental side of EXAFS, making it accessible to non-experts and attracting the largest number of users at synchrotron sources* (Stern 2001 p 51). Unfortunately, over-interpretation, in particular over-estimation of the spatial resolution, is the negative outcome of EXAFS enormous growth among such inexperienced people (*ivi*).

Modern EXAFS, now conceived as, and actually being, an almost unique local structural probe (e.g., Stern 1978; Lee et al. 1981), contributes significantly to understanding the physical behaviours of non-ordered systems such as defective and semi-amorphous materials that have important applications in modern life e.g., catalysts, semiconductors, biomolecules, etc. Perhaps the EXAFS equation may appear to many researchers to be *formidable* (Lytle, 1999 p 131), and yet an ordered sequence of mathematical steps performed by *ad hoc* computer programs allows extracting easily (albeit still by trial and error, occasionally) from the experimentally recorded signal all the intrinsic structural information: a) interatomic distance *R* (down to a 0.01÷0.02 Å); b) coordination number *N* (with a ~10% approximation); c) Debye-Waller factor; d) asymmetry due to thermal expansion. The major drawbacks in the EXAFS state-of-art concern the main approximations non-accounted for by the theory, and consequently neglected (or even ignored) by most users: it is not sensitive to bond angles, and is complex to interpret when the system is too ordered (cf. Gunnella et al. 1990). Moreover, because of the large error inherent in the determination of coordination atoms, the EXAFS analysis fails when the absorber coordination is high: typically for N>8.

Nevertheless, the continuous improvement of EXAFS methods allowed also identifying shake-up and shake-off channels that are associated with multi-electron transition processes (Bernieri and Burattini 1987), and which for a long time had been considered invalid speculations. Much experimental and theoretical effort has been dedicated to determine amplitudes and positions of those excitations and to calculate their relative cross sections (cf. Chaboy et al. 1994, 1995). Although the intensities of such multi-electron excitations are very low, and the unambiguous identification of their energy position and shape is difficult in the presence of the large EXAFS oscillations, their presence affects the EXAFS data, leading to errors in the determination of interatomic distances, coordination numbers and even first-shell anharmonicity (D'Angelo et al. 2004).

**3.2.b. XANES spectroscopy.** The "fine structure" across the absorption edge, although discovered first and interpreted properly on its general lines, has proved to be a rather difficult subject to study in detail. It still has many obscure points, which often make it unfit to be made use of even when powerful computers and modern methods are available. Despite the overall picture is unclear for several reasons, XANES spectra are frequently used in the "fingerprinting" mode e.g., to recognize in the most straightforward way a tetrahedral coordination from an octahedral one (e.g., Mottana et al. 1997), etc. It is physically senseless, although customary, to distinguish between XANES[9] and NEXAFS (cf. Stöhr 1992), as they are the same spectroscopy method. They differ only in that hard and, respectively, soft X-rays are used to scan the sample. In other words, some researchers prefer to use the acronym NEXAFS (Near Edge X-Ray Absorption Fine Structure) to present and discuss X-ray absorption data recorded while doing surface experiments (see later) or the *K*-edges of low Z atoms such as C, N or O.

---

[9] This acronym was used first by Antonio Bianconi (April 1, 1944-) in 1980 while working at SSRL, Menlo Park, USA. It was published by him two years later (Bianconi et al. 1982).

When the absorption effects of atoms impinged by SR started being recorded accurately in their whole extent and complexity, two different approaches to near-edge data presentation, i.e. to XANES, developed: a) a phenomenological - "fingerprinting-like" - description with the simple interpretation of the main features occurring across the main absorption edge, given as $E_0$ (thus from -10 eV to +30 eV with respect to $E_0$); b) an analytical, theoretical study of the MS contributions generated by the photoelectron in relationships to the local geometry of the material probed and to the disorder and defects shown by its atom arrangement around the absorber. The last approach may open up as far as to include the atoms located in high-order coordination shells, in this trying to reach a unified picture of the entire absorption (cf. Stern, 1978; Natoli and Benfatto, 1986). In fact, the two approaches are interconnected; while the first one is typically followed by EXAFS users because it complements their quantitative information on the relative distance among atoms with qualitative data on the coordination and electronic state of the absorber, the second one is followed only by small, theoretically-gifted groups that try deciphering by appropriate calculations the overall physical reactions undergone by the photoelectron displaced from the absorber in the studied material impinged by an electromagnetic radiation of suitable energy. The MS approach, therefore, is essentially a method to calculate the structure of polyatomic matter from first principles, working in the real space (Lee and Pendry 1975; Kutzler et al. 1980; Natoli et al. 1980; Durham et al. 1982; Natoli 1983; Bunker and Stern 1985; Ankudinov et al. 1998; etc.) and searching for an accurate description of the wave function.

Thus, the MS approach proposes and tests algorithms having both theoretical and computational insight. Underneath its application to XANES, there is the assumption that, in the proximity of an edge, the X-ray absorption coefficient depends upon the photoelectron transitions to low-lying unoccupied states in the conduction band of the absorber atom. In the MS framework, both the initial state, which is characterized by a well-defined core level with a localized wave function having precise angular momentum symmetry, and the continuum part, which is represented by modulations of the density of non-occupied electronic states, are calculated. These structures, in fact, are the features near the edge generated by the dipole selection rule, which samples the density of the final states projected onto the angular momentum and convoluted with the core-hole energy width.

In the MS framework, a XANES calculation starts with an approximate molecular potential obtained by partitioning the cluster of atoms considered to be representative of the studied material into distinct atomic and interatomic regions, with each atom enclosed in a sphere of specific radius ("muffin-tin") and an outer sphere enveloping the entire atomic cluster whose size depends by the mean-free-path (typically <10 Å) and the lifetime ($\sim 10^{-15}$ s) of the X-ray excitation. The Coulomb and exchange part of the potential are approximated via the total charge density of the cluster and may or may not consider the existence of dynamical effects. The photoelectron path is made of the several segments that connect the atoms involved in the collision. Calculations are iteratively performed by increasing the size of the cluster and different potentials are considered until reaching calculated spectra that closely resemble the experimental spectrum.

The EXAFS theory had taken about ten years to reach completeness (see above) and from that moment on the EXAFS technique gained wide appreciation among both the scientists and the industrial world. By contrast, although there is general convergence on the MS formalism to explain XANES spectra, a comprehensive, easy to use and universally accepted practical approach to XANES is still lacking, mainly because of the relevant, but still incomplete and somewhat unclear electronic information present in the edge region. At present, there are several formulations, each one based on different algorithms. They underline a gigantic effort at understanding (cf. Natoli et al. 1990, 2003; Rehr and Albers 2000; Ravel and Newville 2005; etc.) and, in addition, reflect the increased computer power that makes these calculations possible in large atomic clusters consisting of up to ~200 atoms or more.

A slow, but steadily increasing theoretical development has taken place ever since the first application of the MS method, which was proposed in the analogy of the Fano's (1935) concept of "shape resonances", which he had conceived for nuclear physics. Indeed, the MS interpretation is its extension to condensed systems. Such "shape resonances" are localized states in the continuum observed in the spectra of diatomic molecules like $N_2$, which could be easily interpreted by the MS theory (Dill and Dehmer 1974; Dehmer and Dill 1975). Only later, it was recognized that there is no true physical reason to make use of different algorithms for XANES and EXAFS, as the MS calculation method may reproduce the entire spectrum, if properly applied (Benfatto et al. 1986).

The first major step ahead in this direction was taken by Calogero Rino Natoli (May 3, 1941-), who first showed that through MS calculations one can determine the absorber to scatterer distance (Natoli 1983). Then, Benfatto et al. (1986) showed how the MS theory applied to liquid solutions connects the lower-energy XANES region to the EXAFS region at high energy via an intermediate energy region where only a few MS contributions contribute to the XAFS spectrum (see also Natoli and Benfatto 1986). Nowadays, it is well recognized that, neglecting the small structures before the edge (i.e., the "pre-edge"), a XAFS spectrum consists of three regions undergoing different scattering modes: a) a full multiple scattering (FMS) region close to the main edge $E_0$ i.e., with -10/20 $\leq E \leq$ +20/25 eV, which is nothing else but the former XAS "fine structure" or "Kossel structure", now called XANES; b) an intermediate multiple scattering (IMS) region (+20/25 $\leq E \leq$ +100/150 eV); and a single scattering (SS) region, which extends to +500 eV or +1000 eV or even more above the edge value $E_0$ and is nothing else but the former "secondary absorption" or "Kronig structure", now called EXAFS (Fig. 4). Within this theoretical framework, the transitional IMS region plays a critical role for the quantitative evaluation of XAFS spectra (Pendry 1983), although it has rarely been investigated in detail so far (e.g., Bugaev et al. 2001; Brigatti et al. 2008). It is now acknowledged that a full XAFS spectrum is the sum of all these regions (Stern 2001), and yet separate mathematical treatments of the XANES and EXAFS regions are still normally done, since the latter one, particularly, is quite simple to apply and is straightforward for most applications, whereas the former requires the cumbersome application of complex computer codes.

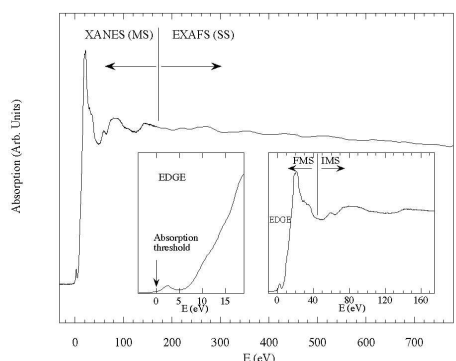

Fig. 4. A modern schematic picture of the XAFS spectrum of the *K*-edge of a transition metal atom showing the MS or XANES region (with in the left inset the FMS sub-region with the pre-edge and the edge rise, and in the right inset the transitional IMS sub-region) and the region of SS or EXAFS (after Mottana, EMU Notes in Mineralogy **6**, 2004 - p. 478, Fig. 4, largely modified).

As a matter of fact, no unified theory nor computer program have been proposed so far that are able to account for the entire spectrum by a single algorithm, although the scenario is wide and in continuous evolution (EXCURVE: Binstead et al. 1991; CONTINUUM: Natoli et al. 1990, 2003; GNXAS: Di Cicco 1995, Filipponi and Di Cicco 2000; FEFF: Newville 2001; Ravel and Newville 2005); MXAN (Benfatto et al. 2001, 2003; WIEN2k: Blaha et al. 2001; SPRKKR: Ebert 2000; and several others optimized for special cases).

The contribution of the upgrade of computing methods (both hardware and software) to XANES development through calculations performed by the MS method cannot be overestimated. Calculations are now possible within both the "muffin-tin" and "non-muffin-tin" approximations (Joly 2001, Hatada et al. 2010), although a major, unfortunately still incompletely worked out problems, first evidenced by Natoli et al. (1986) and studied by Gunnella et al. (1990), lies in the use of potentials (Joly et al. 1999).

### 3.3 Advances in experimental work

**3.3.a. EXAFS**. The algorithms developed in 1974/1975 by Lytle's group (see above) proved their worth in a variety of applications that were previously believed to be possible only by XRD, and are now to be considered classical. They allow data retrieval from experimental recordings performed practically under any condition: at low- as well as high-temperature and pressure (e.g., Itié 1992; Itié et al. 1997; Comez et al. 2002), down to near 0 K and up to 1 Mbar, especially when using the diamond-anvil cell (DAC) in the energy dispersive setting (Bassett et al. 2000). Quick- and Turbo-EXAFS (Dartyge et al. 1986; Pascarelli et al. 1999) provide data in milliseconds that allow live recording reactions while they are in progress, although extracting their structural information then requires some additional time. The polarization of the SR beam has been carefully analysed (Benfatto et al. 1989; Brouder 1990; Pettifer et al. 1990) and made advantage of to measure the atom distribution in materials displaying layered structures (e.g., Manceau et al. 1988; Cibin et al. 2010), and also the distribution and orientation of dopants epitactically growing over certain chemical compounds (e.g., Asakura and Ijima 2001).

Another major step ahead in XAFS due to the power of third-generation synchrotrons is the development of the XAFS-microprobe (SmX: Sutton et al. 1995, 2003; µ-XAS: Mosbah et al. 1999). This apparatus allows recording spectra at micron- and even pixel-size, thus also permitting imaging and quantitative mapping of the atom distribution in the studied materials, and consequently also exposing their zoning and all their intrinsic and extrinsic defects. SmX may take advantage either of the fingerprinting method or, more carefully, of quantitative algorithms. It is particularly interesting for metamorphic petrologists, as it sorts out compositionally different generations of the same mineral, thus allowing a well-planned use of geo-thermobarometers (Dyar et al. 2002; cf. Sutton et al. 2002). A combination of XANES and XRF makes it possible to draw pixel-size single images as well as large maps showing composition and speciation for selected elements (Muñoz et al. 2006).

A field where EXAFS has shown all its worth is in the study of metamict materials (e.g., Greegor et al. 1984; Nakai et al 1987), which are particularly important in the modern world because they are a common by-product of nuclear power plants. Moreover, EXAFS showed its worth also on quasicrystals (Sadoc 1986) i.e., a new category of synthetic materials having non-crystallographic properties that only recently has reached high evidence up to attain the Nobel Prize in Chemistry (2010). Nevertheless, the most exciting results of EXAFS research concern geological materials, particularly those related to the oil and coal industries, and environmental problems produced both by the mining industry and the metallurgical and chemical industrial use of various geological materials. Such studies as determining the size contamination of sediments and mine tailings by As, Pb, U, etc. are to be considered extremely valuable contribution to the integrated study of Man Living Space (cf. Fenter et al. 2002). Indeed, EXAFS is at its best when it studies semi-amorphous ores and ore derivatives, where XRD cannot give information. Moreover, the exact location and ordering of selected trace elements in crystalline hosts (e.g., diamond) and even the measurement of nanometric metal inclusions (e.g., Clausena et al. 1994) are possible by EXAFS, while XRD often cannot even detect their presence. Finally, it is interesting to note that the Continuous Cauchy Wavelet Transform (CCWT) analysis (Munoz et al. 2003), which some time ago seemed to supersede the classical EXAFS treatment because it gives a much wider description of the radial distribution functions and in three dimension, in contrast with the classical method which is limited to a one-dimensional representation, has declined in the general appreciation: the classical method is too well rooted among users to be easily replaced.

From the very first stages of SR extraction from storage rings, EXAFS was optimized also for surface experiments. This special technique, called SEXAFS (Surface Extended X-ray Absorption Fine Structure), concerns the study of electronic transitions from core levels of atoms located at the surface of solids. The peculiar local character of core level excitations in the X-ray absorption processes makes it attractive to study the atoms at the surface of solids and to investigate both their local structure and localized electronic states. The knowledge of the atomic arrangement of neighbour atoms around a selected atom on a surface is important in many industrial studies such as chemisorption, oxidation processes and catalysis. SEXAFS has gained wide appreciation because the local structure of chemisorption sites and the local structure of surface amorphous oxides are basic information that are not directly given by any other technique based on diffraction methods e.g., Low Energy Electron Diffraction (LEED).

SEXAFS was born in the middle of seventies as soon as the tunable and intense SR X-rays of storage rings become available. Indeed, in order to efficiently start, it had high flux requirements, owing to the very low concentration of surface atoms: the main experimental problem of all surface X-ray spectroscopies was and still is how to enhance the sensitivity at surface. First Lukirskii and Brytov (1964), using the continuum bremsstrahlung of a standard X-ray tube, and later Gudat and Kunz (1972), using SR, demonstrated that the total electron yield (TY) by the sample surface is proportional to the bulk absorption coefficient, while having a sampling depth of the order of magnitude of few tens of Å. Therefore, TY is fit for surface investigations, but it could not be proficiently used till a strong enough SR would come in. The first SEXAFS experiment having high surface sensitivity was carried out at the Stanford Synchrotron Radiation Facility (SSRF, now SSRL) by detecting the Auger electron yield (Bianconi et al. 1977). The absorption spectrum of the surface Al atoms in the top monolayer of an aluminium crystal could be distinguished from the aluminium bulk spectrum: contrast was increased by selecting an Auger line originating from the Al atoms at the surface, which interacted with chemisorbed oxygen producing an inter-atomic, Al-O Auger transition. Both the total electron yield (Citrin et al. 1978) and the Auger electron yield (Stöhr et al. 1978) techniques were then made use of to measure the signal of different atomic species chemisorbed on solid materials. These experiments brought with them, as an immediate consequence, that the relaxation undergone by the atoms in the exposed substrate started being studied too (Bianconi and Bachrach 1979; Bianconi et al. 1979).

As shown above, SEXAFS concerns the study of the absorption coefficient modulations over a range of photoelectron wave-vectors above ~3 Å$^{-1}$. In this range, the experimental data can be analyzed in the framework of the well-established EXAFS theory (Stöhr 1986; Norman 1986), and contribute to policy-making particularly on matters related to environmental remediation (e.g., Brown et al. 1999; Sherman and Randall 2003). By contrast, the XAFS structures present in the low-energy range (few tens of eV), which contain information on the local geometry i.e., on bond distances and bond angles, should be evaluated and discussed in the framework of the XANES theory (Bianconi and Marcelli 1992).

**3.3.b. XANES**. In the near-edge fine structures of XANES, the higher-order terms of the correlation function of the atomic distribution become much more important than the pair correlation function of the atomic distribution probed in the EXAFS range. They too have been intensively studied using SR ever since the first application of this as the X-ray source. However, due to the lack of a reliable theoretical method of analysis, early XANES studies on local geometry determination have been limited to a fingerprint approach using model compounds. This compelled scientists to perform a preliminary step that was considered useful, i.e., to re-record, using SR as the source and updated instrumental detection setups, most if not all the near-edge XAS features of the many materials that had been collected during the previous fifty years. As pointed out before, many of them were actually very accurate, and yet they all underwent confirmation.

Among the best results on this line of work, those on Mn (Belli et al. 1980), Fe (Waychunas et al. 1983), and Ti (Waychunas 1987; Paris et al. 1993; Farges et al. 1996) are worth mentioning: they all concern transition-metal atoms of high significance for industry. These studies also updated certain already well-known XANES properties such as, e.g., the chemical shift, or angular dependence, or coordination dependence (cf. Mottana, in preparation). Moreover, a completely new emphasis was given to the role of the pre-edge features, since instrumental upgrade made it possible to record them with resolution and intensity much higher than ever done before. (Wu et al., 2004) Important advances towards a quantitative theoretical analysis of the surface XANES data have also been carried out (Bianconi and Marcelli 1992). In particular, priority has been given to biomolecules (e.g., Shulman et al. 1976; Bianconi 1983), to soft semi-amorphous compositionally inhomogeneous compounds (e.g., bitumen and coal: Wong et al. 1983; catalysts), and to glassy materials (e.g., Greaves et al. 1984; Greaves 1985), because their very low crystallinity hinders studies using XRD, while allowing XAFS, which does not requires LRO for detection and study.

Practically, two different lines of approach to XANES have developed, which both persist up to now. The first and easiest one makes use of the fingerprinting method and extracts from the XANES region the electronic information required to best interpreting EXAFS quantitative results, e.g., the oxidation state of the absorber. Typical examples of this trend are the many studies concerning iron and other transition atoms, starting with the first realization that energy and shape of the pre-edges of transition-atom-bearing minerals could bear quantitative information of the relative amounts of e.g., coexisting $Fe^{2+}$ and $Fe^{3+}$ (Bajt et al. 1994). Furthermore, it was shown that the intensity of the pre-edge brings information on the relative coordination e.g., of $Ti^{[4]}$ and $Ti^{[6]}$ (Paris et al. 1993). This trend has been pursued particularly by geochemists and is in general universally acknowledged for transition metals and metalloids, its best example of application being on the iron-bearing natural materials and on the oxidation state of arsenic in waters and soils. Indeed, studies on pre-edge energy and shape have shown that these two features account for both coordination and oxidation state of Fe in known minerals (e.g., Wilke et al. 2001; Berry et al. 2008) and may foresee conditions that are not yet realized on Earth, but which might occur on the Moon and in meteorites. Similarly, Sherman and Randall (2003), although using a somewhat crude fingerprinting approach, could demonstrate how reacting $As^{5+}$-bearing solutions with a $Fe^{3+}$-bearing oxide- and hydroxide-rich substrate can contribute to reclaiming heavily polluted environments, essentially by reducing poisonous $As^{5+}$ to almost harmless $As^{3+}$. Environmental studies benefit the most of XANES because this kind of spectroscopy is utterly independent on the state of the absorber, while being able to reveal very low contents of it (ppb)(Brigatti et al., 2000). Thus, recently, several manuscripts appeared that deal with aerosols and nanoparticles suspended as fine dust in the low layers of the atmosphere, or even permanently trapped in continental ice, such as in Antarctica (Lama et al. 2012; Marcelli et al. 2012; Schlaf et al. 2012; etc.).

Chemical reactions also occur at the interface between two solid layers of different composition and/or structure, which often are packed together to form multilayers. These have lately become a kind of artificial material highly interesting for advanced industry and in many scientific applications. As a matter of fact, for multilayers, the use of XAFS in the total reflection mode (ReflEXAFS: Davoli et al. 2003) gives results that are far more useful than SEXAFS, especially for samples that are grown by molecular beam epitaxy, the added layers of which have nanometric thickness (López-Flores et al. 2009).

Another major step ahead in XAFS due to third-generation synchrotrons is the further upgrade of XAFS-microprobes, which allow recording spectra at micron- and smaller scales limited only by the pixel size of the available detectors. They permit imaging and quantitative mapping of a sample showing intrinsic and extrinsic defects and compositional variations. Indeed, for all applications from biology to environmental science it is always more important to fully understand the concentration profile and the chemical state of the major number of elements present in a given material. In the last decade, several two-dimensional (2D) X-ray imaging methods have been developed for *in situ* analysis of large sample areas. These data show the distribution of particles in a sample, while XAFS reveals the kind of elements present and their chemical state. Moreover, X-ray microscopy and tomography can be also combined with XAFS resulting in a much more effective analytical technique. A three-dimensional (3D) computed tomographic XANES approach has been recently demonstrated (Meirer et al. 2011). It uses a CCD detector having a space resolution of a few μm. The ability to simultaneously probe morphology and phase distribution in complex systems at multiple length scales may unravel the interplay of nano- and micrometre-scale factors that are at the origin of a material macroscopic behaviour. The newest X-ray layouts combine full-field transmission X-ray microscopy (TXM) with XANES spectroscopy, to follow 2D and 3D morphological and chemical changes with a resolution of tens of nanometres in thick samples, allowing the analysis of large areas over a time from few min to few hours.

This mass of new information prompted the in-depth theoretical studies on XANES outlined previously and the formulation of the new software based on the MS theory extensively described there. It is now customary, even among industrial users, to accompany their experimental results with structural (or pseudo-structural) interpretations based on theoretical simulations performed using anyone of the software packages listed above, most of which are freely available. Therefore, XAFS in all its various techniques of application is now one of the most specialized techniques of largest industrial interest. Modern industry – particularly chip technology, which is essential for both computers and communication tools – needs to know as precisely as possible the state of the surface of the material and of interfaces rather than its bulk. In fact, most reactions involved in technological operations (e.g., lithography of circuits; epitaxial growth of either required dopants and of unwanted impurities, these ones to be removed) occur at the surface or in the first few layers of the material, and are further complicated by the relaxation effect any well-ordered bulk structure undergoes when it is suddenly interrupted. Consequently, both fingerprinting and MS techniques typical of SEXAFS and ReflEXAFS have been used to cope with these assignments on a variety of materials, starting from ores up to proteins, in a practically continuous, numberless sequence.

## 3 Conclusion

XAFS i.e., X-ray absorption fine spectroscopy (now intended to be XANES+EXAFS, performed together at high resolution), is certainly one of the most advanced techniques of material investigation particularly among industrial and applied scientists. Indeed, SR facilities are flooded by requests of users willing to apply XAFS methods to different materials and phenomena. (Fukui et al., 2001) XAFS-related methods such as SEXAFS (Surface EXAFS) and ReflEXAFS (Reflectivity EXAFS), or others too recent or too poorly pushed to be mentioned here such as e.g., XES (X-Ray Emission Spectroscopy), RIXS (Resonant Inelastic X-Ray Scattering), TXRF (Total Reflection X-ray Fluorescence), XMCD (X-ray

Magnetic Circular Dichroism), and DAFS (Diffraction Anomalous Fine Structure), are successfully applied to complex systems, the deep knowledge of which is requested by modern research aiming to industrial applications. New instruments are available, which allow performing polarized experiments on single crystals and, recently, even µ-XAFS, i.e., XAFS performed on micrometre-size spots or microparticles, is increasingly developed, thus preparing the further evolution towards nanoparticle studies, which still mostly rely on high-resolution transmission X-ray microscopy (TXRM) with a spatial resolution in the range ~30-50 nm. Nowadays, time-resolved experiments using either dispersive or QUICK-XAFS layouts permit following the advance of physical phenomena or of chemical reactions with a time resolution from seconds down to few tens of microseconds.

The continuous growth of XAFS documented here pays a particular attention to new ideas and their theoretical extension. Early XAS contributed to the understanding of the atomic theory of matter and to the establishing of Materials Science (cf. Mottana, in preparation). Since 1975, XAFS contributed to its advanced understanding; starting from the characterization of complex or extremely diluted materials up to the interpretation of order-disorder phenomena under the peculiar aspect of electronic interactions among neighbouring atoms in a complex chemical environment. The explosion of modern XAFS opened new fields and generated unexpected results. All this historical sequence of projects, tests, and events brings support to the opinion, widespread among the scientific community, that much greater results are to be expected, such as those offered in the near future by the incoming availability of fourth-generation SR coherent X-ray sources.

However, now, the front line of XAFS research is at trying to make the best out of FEL (Free-Electron Laser). This novel, extremely sophisticated energy source, allows studies on the near-edge fine structure of solids by means of femtosecond soft X-ray pulses. So far, only pioneering attempts on model materials have been successfully carried out (Bernstein et al. 2009). Such preliminary experiments could be carried out only under many strong constraints on experimental conditions such as sample geometry, spectrographic energy dispersion, single shot position-sensitive detection. In addition, they used a data normalization procedure that eliminates the severe fluctuations of the incident intensity in space and in photon energy present in FEL, which resemble those occurring in the very first synchro-trons. Nevertheless, all these pioneering works gave convincing results, thus opening new potential applications to XAFS that will meet with the increasing demands of Materials Science as the relevant branch of Physics and, without any doubt, of the most advanced, high-tech industry, which is definitively oriented to make the best out of its nanometre-size products.

**Acknowledgments**. Our attempt at sketching the development of XAS and at clarifying more extensively its transition to XAFS up to the present time is based on more than two decades of readings and discussions with many people, whom we thank greatly for contributing to our effort. Among them, we sort out Antonio Bianconi, Maurizio Benfatto, Jesús Chaboy Nalda, Giannantonio Cibin, Ivan Davoli, Rino Natoli, Hiroyuki Oyanagi, Eleonora Paris, Piero Pianetta, Trevor Anthony Tyson, and Ziyu Wu. Nevertheless, the responsibility of this historical account, which necessarily involves selection and evaluation of persons and works, lies entirely upon us. We certainly forgot relating about some important research papers, but both the capacity of our memory and the length allowed to this contribution have limitations. We sincerely apologize to all colleagues whose worthy papers we neglected to quote.

**Analytical index**

Lytle, Farrel W.

Many-body correlation function

Material Science

MBE

Mean free path

Metamict

Microprobe

Model compound

Model, one-dimensional

Model, three-dimensional

Modulation

Molecular beam epitaxy

Monochromator, bent-crystal

Monochromator, curved-crystal

Monochromator, double-crystal

Monochromator, flat-crystal

Moseley, Henry Gwinne J.

MS

*M*-series

Multichannel correlation function

Multilayer

Multiple transition theory

MXAN

µ-XAS

Natoli, Calogero Rino

Near-edge structure

TXRF

TY

Undulator

Van Nostrand, Robert G.

Wave function

Wave, scattered

Wave, spherical

Wavelength

White line

Wiggler magnet

XAFS

XANES

XMCD

X-ray absorption spectrum

X-ray discovery

X-ray emission

X-ray polarization

X-ray source

X-ray spectrum, characteristic

X-ray spectrum, continuum

X-rays, hard

X-rays, monochromatic

X-rays, polychromatic

X-rays, soft

XRD

XRF

Z